\documentclass{aa}
\usepackage{natbib}\bibpunct{(}{)}{;}{a}{}{,}
\usepackage{graphicx}
\usepackage{hyperref}
\begin{document}

\title{Acceleration of GRB outflows by Poynting flux dissipation}

\author{G. Drenkhahn\thanks{e-mail: \texttt{georg@mpa-garching.mpg.de}}} 

\institute{Max-Planck-Institut f\"ur Astrophysik, Postfach~1317,
  85741~Garching bei M\"unchen, Germany}

\date{A\&A 387,714 --- Received 21 December 2001 / Accepted 11 March 2002}

\abstract{We study magnetically powered relativistic outflows in which
  a part of the magnetic energy is dissipated internally by
  reconnection.  For GRB parameters, and assuming that the
  reconnection speed scales with the Alfv\'en speed, significant
  dissipation can take place both inside and outside the photosphere
  of the flow.  The process leads to a steady increase of the flow
  Lorentz factor with radius.  With an analytic model we show how the
  efficiency of this process depends on GRB parameters.  Estimates are
  given for the thermal and non-thermal radiation expected to be
  emitted from the photosphere and the optically thin part of the flow
  respectively.
  
  A critical parameter of the model is the ratio of Poynting flux to
  kinetic energy flux at some initial radius of the flow.  For a large
  value ($\ga 100$) the non-thermal radiation dominates over the
  thermal component.  If the ratio is small ($\la 40$) only prompt
  thermal emission is expected which can be identified with X-ray
  flashes.
  
  }

\maketitle

\defcitealias{spruit:01}{Paper~I}

\section{Introduction}
\label{sec:introduction}

To overcome the compactness problem of $\gamma$-ray bursts (GRBs)
\citep[e.g.][]{piran:99} the central engines must produce radiating
material moving ultra-relativistically fast towards the observer.  GRB
models must therefore describe an energy source which not only
releases energy of around $10^{52}\,\mathrm{erg/sterad}$ but must also
explain the `clean' form of the energy.  To produce the high Lorentz
factors of the order of $10^2$--$10^3$
\citep{fenimore:93,woods:95,lithwick:01} which are needed only a small
fraction of the total energy can exist in form of rest mass energy of
the matter involved.

The popular models involving compact objects or the collapse of a
massive star to a black hole must include mechanisms how the energy is
transported into a space region with few baryons.  Otherwise large
amounts of mass are expelled which cannot be accelerated to high
Lorentz factors.  The initially released energy could leave the
central polluted region by neutrinos which annihilate to a pair plasma
further away \citep{berezinskii:87,goodman:87,ruffert:97}.  But due to
the small cross section of neutrinos the efficiency is low and most of
the energy escapes as neutrinos.

A Poynting flux dominating outflow will naturally occur if the compact
object rotates and possesses a magnetic field.  The luminosity will be
fed by the rotational energy reservoir of the central object.  Models
involving an magnetised torus around a black hole \citep{meszaros:97}
or a highly magnetised millisecond pulsar
\citep{usov:92,kluzniak:98,spruit:99} would produce such a
rotationally driven Poynting flux.  Extraction of energy from the
central object by this magnetic process is potentially very efficient
and fast.

In order to obtain not only a large energy extraction but also the
observed large bulk Lorentz factors, the Poynting flux must be
converted to kinetic energy.  The simplest available magnetic
acceleration models, in which the flow is approximated as radial, are
problematic in this respect.  In the classic non-relativistic case
\citep{weber:67,belcher:76} a dominating initial Poynting flux can
transfer $1/3$ of its energy to the matter.  If the flow is initially
relativistic however almost no acceleration is possible
\citep{michel:69}.  The physical reason lies in the singular field and
flow geometry of a purely radial flow.  In this case the magnetic
pressure gradient balances the magnetic tension force and no
acceleration occurs.  An imbalance between the pressure gradient and
the tension force occurs in non-radial outflows, if the flow lines
diverge faster with radius than in the radial case
\citep{begelman:94,takahashi:98}.  Detailed 1-dimensional calculations
have been made which show how such a flow divergence can come about
\citep{beskin:97,daigne:01}.

In this paper we show that there is a second process which naturally
leads to efficient conversion of Poynting flux to bulk kinetic energy.
If the magnetic field in the outflow contains changes of direction on
sufficiently small scales, (a part of) the magnetic energy is `free
energy' which can be released locally in the flow by `fast
reconnection' processes.  Such a decay of magnetic energy, if it can
occur rapidly enough, has two desirable effects.  First it provides a
source of energy outside the photosphere which is converted directly
into radiation, without the relatively inefficient intermediate step
of internal shocks \citep[ hereafter
\citetalias{spruit:01}]{spruit:01}.  Secondly, it leads to an outward
decrease of magnetic pressure, which causes a strong acceleration of
the flow and conversion of Poynting flux to kinetic energy.  In the
present work, we concentrate on the acceleration effect, and show how
it depends on the parameters (energy flux, baryon loading) of a GRB.
This aspect of the model can be illustrated with analytic
calculations.  In a future paper we show, with more detailed numerical
results, how the dissipated magnetic energy can also power the
observed prompt radiation.

Changes of direction of field lines must occur in the flow in order
for energy release by reconnection to be possible.  These can occur
naturally in a number of ways.  If the magnetic field of a rotating
central object is \emph{non-axisymmetric} the azimuthal part of the
magnetic field in the flow changes direction on a length scale
$\lambda\approx \pi v/\Omega$, where $v$ is the flow velocity and
$\Omega$ the angular frequency.  For an inclined dipole this yields
the `striped' field in pulsar wind model of \citet{coroniti:90} where
magnetic energy is released by the annihilation of the antiparallel
field components.  Field decay by reconnection was applied to pulsar
winds \citep{coroniti:90,lyubarsky:01} and also to GRBs
(\citealp{thompson:94}; \citetalias{spruit:01}).

In this paper we investigate the dynamics of a magnetically powered
outflow in which some of the energy dissipates by reconnection.  With
the assumption that the flow is highly dominated by magnetic energy
and that the thermal energy is negligible we derive the velocity
profile of the flow.  The results provide estimates of the Lorentz
factor of the flow, the photospheric radius, and the amount of energy
that can be converted into non-thermal radiation.  We investigate
under which conditions prompt emission is expected and whether a
considerable amount of thermal radiation can be produced.  These
predictions can then be tested against observations of the thermal
component in GRB spectra \citep{preece:00}.

\section{Model description}
\label{sec:model}

Highly magnetised spinning compact objects, e.g.\ millisecond pulsars
or tori around black holes, are sources of Poynting flux that can
power GRBs.  They produce a plasma-loaded electromagnetic wind
travelling outward and are fed by the rotational energy of the central
object.  In the wind of an aligned rotator the magnetic field is
ordered and stationary.  If ideal MHD applies, and the wind is radial
in the poloidal plane, a large fraction of the total luminosity is
bound to stay in form of Poynting flux.  The picture changes in the
case of an inclined rotator or any other source producing a
non-axisymmetric rotating magnetic field.  If the emitted Poynting
flux contains modulations of the field it also carries along free
magnetic energy, which can be extracted by reconnection processes.  In
these processes the field rearranges itself to a energetically
preferred configuration while the energy released is transfered to the
matter.  Because perfect alignment of magnetic and rotation axis is a
special case it is likely that most astrophysical objects produce
modulated Poynting fluxes containing free magnetic energy.

A necessary condition for the existence of free magnetic energy in
the flow is the field variation on small scales.  For reconnection
processes differently oriented field lines must come close to each
other.  Therefore the length scale on which the orientation of
magnetic field lines change controls the speed of the field
dissipation.  The smaller the length scale is the faster the field can
decay.

The general large scale magnetic field structure expected to be
produced by a rotating object was discussed in \citetalias{spruit:01}.
It is useful to consider simplified flow geometries along the
equatorial plane and along the rotation axis as examples.  In the
equatorial plane an inclined rotator will produce a `striped' wind
\citep{coroniti:90}.  It consists of an electromagnetic wave in which
the azimuthal field component varies with a wave length of $2\pi
c/\Omega$.  Along the rotation axis the wave will have a circular
component with the same wave length.  Such wave-like field variations
are present in general if a non-axisymmetric magnetic field component
is present.  The equatorial plane of an inclined rotator is only a
prototype to illustrate the field geometry.  In general, wave-like
variations occur at all latitudes.  If the rotator is aligned the
field will be axisymmetric.  Then, the magnetic field geometry looks
like a wound up spiral on all cones of equal latitude.  This field
geometry is present in case of a jet-like outflow.  Here, the magnetic
field does not vary on small scales along the outflow direction.  The
differently directed field components lie on opposing sides of the
rotation axis.  In the context of a jet-like outflow the typical
length scale of the field variation is the diameter of the jet cone
$r\vartheta$ where $\vartheta$ is the jet opening angle.

In both of these field geometries MHD instabilities can promote
reconnection processes.  For wave-like variations current sheets form
and tearing instability will lead to reconnection.  For a polar
jet-like outflow of an aligned rotator the field configuration is
highly unstable to the kink instability \citep[e.g.][ see also
\citetalias{spruit:01}]{bateman:80}.  It is plausible that the kink
instability working in this case will distort the geometry after some
time so that also wave-like variations come into play.  This leads to
non-periodic and highly irregular waves.  Our model assumes the
longitudinal field variation to be periodic so that the complicated
effects of any non-periodicity is neglected.  We consider both
limiting cases for the small scale field variations though wave-like
structures seem to be more general.

Near the source the flow is accelerated magnetocentrifugally (and
perhaps thermally).  It will be accelerated up to a distance around
the Alfv\'en radius and then start to become radial asymptotically.
The poloidal and azimuthal field components at the Alfv\'en radius are
similar in magnitude.  Beyond this point their ratio scales as
$B_\phi/B_r\sim r$, so that the radial component soon becomes
negligible at a couple of Alfv\'en radii.  The Alfv\'en point lies
always inside the light radius $c/\Omega$ and if the magnetic field
dominates, like in our case, the Alfv\'en radius and light radius are
almost equal.  Thus we can simplify the flow and field geometry at
source distances $r > \mbox{a~few}\times c/\Omega \approx
10^{7}\,\mathrm{cm}$ by assuming a purely radial flow with an
azimuthal magnetic field.  At this distance gravity effects can also
be neglected.  The magnetocentrifugal effects accelerate the flow to
fast magneto-sonic velocity.  Because we work in the cold limit (see
below) and approximate the magnetic field to be purely poloidal the
magneto-sonic velocity is equal to the Alfv\'en velocity.  The initial
flow velocity is set to the Alfv\'en velocity at some initial radius
$r_0\ga c/\Omega$.

To make a simple approach feasible analytically we have to make
further approximations.  The flow is treated stationary and its
thermal energy is neglected (`cold' limit).  This `cold' approximation
is quite good in the optically thick region since no energy can be
lost by radiation anyway.  All of the dissipated energy is always
converted into kinetic form.  If the flow is optically thin the
radiation produced by dissipation will freely escape and this energy
part will not be converted into kinetic energy.  Our model
overestimates the kinetic energy gained in the optically thin regime.
Statements about the radius of the photosphere, where the flow changes
from optically thick to thin, or the Lorentz factor there will hold
rather robustly.

\subsection{Magnetic field dissipation}
\label{sec:dissi}

The dissipation is modelled by using the typical length scale of the
magnetic field $\lambda$ and a fraction of the Alfv\'enic velocity
$\epsilon v_\mathrm{A}$ where $\epsilon$ is an dimensionless factor.
The idea is that the magnetic field lines with different directions
get advected towards a reconnection centre where the field dissipates
\citep{petschek:64}.  This advection happens with a fraction of
$v_\mathrm{A}$.  The decay of mean field must also depend on
$v_\mathrm{A}$ in a similar fashion.  Though there are elaborate
models on the reconnection physics
\citep[e.g.][]{coroniti:90,thompson:94} we prefer to express this
rather uncertain topic in form of the dimensionless free parameter
$\epsilon$ and the Alfv\'enic speed $v_\mathrm{A}$.  All the uncertain
physics in this picture is taken up by $\epsilon$.  Our ansatz for the
time scale of the mean field decay, in a comoving frame is then
\begin{equation}
  \label{eq:tau}
  \tau_\mathrm{co}
  = \frac{\lambda_\mathrm{co}}{\epsilon v_\mathrm{A,co}}
\end{equation}
where quantities considered in the comoving frame are indexed with
`co'.  This comoving frame moves with the mean large scale bulk flow
motion so that the small scale motion is neglected.

The reconnection takes place at certain reconnection centres in the
flow.  The typical distance between these reconnection centres also
influence the rate of the overall field dissipation.  Since we regard
the field dissipation in all generality and for a variety of field
geometries we cannot explicitely model these small scale details about
the density of reconnection centres.  This uncertain issue must also
be handled by the free parameter $\epsilon$ so that $\epsilon$
controls the \emph{average} field dissipation on larger length scales.
  
At first sight $\epsilon<1$ is an upper limit since for $\epsilon=1$
reconnection would happen everywhere with an advection speed of $c$.
If the advection towards the reconnection centres happens with almost
$c$ large current densities are required.  The MHD condition might
break down leading to an additional decay of magnetic field.  This
effect could be parameterised by an larger value of $\epsilon$ so that
an upper limit of 1 may not be strict.

For most of the paper we will work with a fiducial value of
$\epsilon=0.1$.  One should keep in mind that $\epsilon$ is perhaps
the most uncertain quantity of the model because it may not be
constant and its value cannot be estimated by first principles in
general.

The length scale for the dissipation $\lambda_\mathrm{co}$ depends on
the nature of the outflow as discussed in the last section.  We will
distinguish the two cases where the field variation is encountered on
length scales longitudinal to the flow direction (called
\emph{longitudinal case} in the rest of the paper) and where this
length scale is transversal to the flow direction (\emph{transversal
  case}).  The transversal case is found in a polar outflow of a
aligned rotator where the field components having different directions
lie on opposite sides of the rotation axis.  For mathematical
simplicity we will regard here the two limiting cases only and make a
few notes on the mixed case later in this study.

The longitudinal and the transversal length scales
$\lambda_\mathrm{lo}$, $\lambda_\mathrm{tr}$ in the comoving frame
scale differently with the Lorentz factor of the flow $\Gamma$:
\begin{eqnarray}
  \label{eq:lambda_equ}
  \lambda_\mathrm{lo,co}
  &=& \frac{2\pi c \Gamma}{\Omega}
  \ ,\\
  \label{eq:lambda_jet}
  \lambda_\mathrm{tr,co}
  &=& \vartheta r
\end{eqnarray}
where $\vartheta$ is some kind of an opening angle of the polar
outflow.  $\lambda_\mathrm{tr,co}$ does not scales with $\Gamma$
because it denotes a length which is perpendicular to the direction of
motion being the same in the lab and comoving frame.

\section{The dynamics of the flow}

The reconnection processes described in the last section will change
constantly the structure of the magnetic field on a length scale of
the order of the wave length (small length scale of the problem).
Though, the azimuthal field line stretching will keep the field
aligned predominantly perpendicular to the flow direction.  The exact
field structure is not important because only the magnetic energy
density $B^2/(8\pi)$ enters the dynamic equations.  In the following,
$\vec B$ denotes the dynamical effective transversal magnetic field
which is constant over small scales.  The induction equation will
still be valid for the effective field.

\subsection{Conservation laws}
\label{sec:math}

The dynamics of the flow is governed by the ideal MHD equations for
the conservation of energy, momentum and mass.  For the relativistic
treatment the equations are formally best written in tensorial form
\citep[e.g.][]{bekenstein:78}:
\begin{equation}
  \label{eq:enmom_tens}
  {T^{\mu\nu}}_{;\nu} = 0
  \ ,
\end{equation}
\begin{equation}
  \label{eq:cont_tens}
  (\rho u^\mu)_{;\mu} = 0
\end{equation}
where 
\begin{eqnarray}
  T^{\mu\nu}
  &=& \underbrace{w u^\mu u^\nu + p g^{\mu\nu}}_{\mbox{matter part}}
  \nonumber\\
  &&+ \underbrace{\frac{1}{4\pi} \left[
      \left(u^\mu u^\nu + \frac{1}{2} g^{\mu\nu} \right)
      b_\alpha b^\alpha - b^\mu b^\nu
    \right]}_{\mbox{electromagnetic part}}  
\end{eqnarray}
is the energy-momentum tensor for ideal MHD.  The signature $(-+++)$
is used for the metric tensor $g^{\mu\nu}$.  Here, $\rho$, $w$ and $p$
are the mass density, the enthalpy density and the pressure in the
proper frame of the fluid moving with a 4-velocity $u^\mu=(\Gamma,\vec
u)$.  $b^\mu={}^*F^{\mu\nu} u_\nu$ is the the 4-vector of the magnetic
field where ${}^*F^{\mu\nu}$ is the dual electromagnetic field
strength tensor.

Now we choose a spherical coordinate system centred on the central
engine.  The flow is assumed to be spherically symmetric and the field
dominated by its toroidal component.  In this case $\vec u\perp\vec B$
and the components of the magnetic four vector are simply
$b^\mu=(0,-\vec B/\Gamma)$ and $b_\mu b^\mu = B_\mathrm{co}^2 =
B^2/\Gamma^2$.  Writing (\ref{eq:enmom_tens}) and (\ref{eq:cont_tens})
in coordinate form and assuming stationarity gives the conservation
laws for energy, momentum and mass
\begin{eqnarray}
  \label{eq:en_cons}
  0 &=& 
  \partial_r r^2 \left(
    w\Gamma u + \frac{\beta B^2}{4\pi}
  \right)
  \ ,\\
  \label{eq:mom_cons}
  0 &=& \partial_r r^2 \left(
    w u^2 + \left(1+\beta^2\right) \frac{B^2}{8\pi}
  \right) + r^2 \partial_r p
  \ ,\\
  0 &=&
  \partial_r r^2 \rho u
\end{eqnarray}
where $\beta=u/\Gamma$ \citep{koenigl:01,lyutikov:01}.  By integrating
the mass and energy equations one obtains the total mass loss per time
per sterad
\begin{equation}
  \label{eq:Mdot0}
  \dot M = r^2 \rho uc  
\end{equation}
and the total luminosity per sterad
\begin{equation}
  \label{eq:L}
  L = r^2 \left(
    w \Gamma u c + \frac{\beta B^2}{4\pi}c
  \right)
  \ .
\end{equation}

The enthalpy density $w$ includes the rest mass energy density $\rho
c^2$.  In the following we will assume a cold flow with $w=\rho c^2$,
$p=0$.  Then, the momentum equation can be integrated and the
conservation laws read
\begin{eqnarray}
  \label{eq:m_cons}
  \dot M &=& r^2 \rho u c
  \ ,\\
  \label{eq:e_cons}
  L &=& \Gamma \dot M c^2
  + \beta c \frac{(rB)^2}{4\pi}
  \ ,\\
  \label{eq:p_cons}
  \mbox{const.} &=&
  u \dot M c + (1+\beta^2) \frac{(rB)^2}{8\pi}
  \ .
\end{eqnarray}
In the energy equation (\ref{eq:e_cons}) one can identify the kinetic
energy flux per sterad $L_\mathrm{kin}=\Gamma \dot M c^2$ and the
Poynting luminosity per sterad $L_\mathrm{pf}=\beta c (rB)^2/(4\pi)$.

Taking the flow to be cold and eliminating $(rB)^2$ from
(\ref{eq:e_cons}), (\ref{eq:p_cons}) shows that $u$ is a constant
function of $r$.  Finding an exact accelerating solution of the energy
and momentum equations without thermal pressure is not possible.  By
using an evolution equation for the magnetic field $B$ (see
Sect.~\ref{sec:evomag} below) and combining it with the energy
equation (\ref{eq:e_cons}) one obtains an accelerating solution but
violates the momentum conservation.  Luckily, the error made by that
becomes small in the ultra-relativistic limit.  Then, $\Gamma\approx
u$, $\beta\approx 1$ so that the first term in the momentum
Eq.~(\ref{eq:mom_cons}) becomes small since it approaches the form of
the rhs of the energy Eq.~(\ref{eq:en_cons}).  As a consequence the
thermal pressure gradient term of (\ref{eq:mom_cons}) must also be
small.  Setting the pressure to zero and solving the energy equation
means that the momentum equation is \emph{almost} satisfied.  Since we
only consider ultra-relativistic flows the error made is small which
justifies the use of the cold approximation.

\subsection{A note on the ideal MHD approximation}

In the treatment above the ideal MHD approximation was used.  But a
key ingredient of the model is the existence of field dissipation for
which ideal MHD is not applicable at first sight.  The field
dissipation acts like an effective diffusivity in the plasma so that
the effective mean electric field in a comoving frame does not vanish.
Since a substantial electric field $\vec E_\mathrm{co}$ would
contribute to the comoving energy density, the question arises if it
can be neglected.  We found from a more detailed numerical
investigation (in preparation) that the comoving electric field is in
fact small, and we use this advance knowledge to neglect its
contribution to the dynamics here.

\subsection{The initial conditions of the flow}

Let $\sigma$ be the ratio of Poynting flux to matter energy flux:
\begin{equation}
  \label{eq:sigma}
  \sigma 
  = \frac{L_\mathrm{pf}}{L_\mathrm{kin}}
  = \frac{\beta (rB)^2}{4\pi \Gamma\dot M c}
  = \frac{B^2}{4\pi\Gamma^2\rho c^2}
  \ .
\end{equation}
$\sigma$ is also the magnetisation parameter of the plasma, describing
the ratio of the proper magnetic energy density to the proper energy
density of the matter.  The Alfv\'en 4-velocity in the comoving frame
is
\begin{equation}
  \label{eq:alfspeed}
  u_\mathrm{A}
  = \frac{B/\Gamma}{\sqrt{4\pi \rho c^2}}
  = \sqrt{\sigma}
\end{equation}
with the regular dimensional velocity counterpart
\begin{equation}
  v_\mathrm{A}
  = c \frac{u_\mathrm{A}}{\sqrt{1+u_\mathrm{A}^2}}
  = c \sqrt{\frac{\sigma}{1+\sigma}}
  \ . 
\end{equation}
At an initial radius $r_0$, where the flow starts with the Alfv\'en
speed (discussed in Sect.~\ref{sec:model}), the relation between
initial 4-velocity $u_0$ and initial Poynting flux ratio $\sigma_0$ is
simply
\begin{equation}
  \label{eq:u0}
  u_0 = u(r_0) = u_\mathrm{A}(r_0) = \sqrt{\sigma_0}
  \ .
\end{equation}
The total energy and the mass flux are linked by
\begin{equation}
  L = \left(\sigma_0 + 1\right) \Gamma_0 \dot M c^2
\end{equation}
so that $\dot M$ can be expressed in terms of $L$ and $\sigma_0$:
\begin{equation}
  \label{eq:Mdot}
  \dot M = \frac{L}{c^2 (\sigma_0+1)^{3/2}}
  \ .
\end{equation}

In the GRB case the flow must start highly Poynting flux dominated
with $\sigma_0\ga100$ so that $u\gg1$ at all distances.  One can
therefore set $\sigma_0+1\approx\sigma_0$, $u\approx\Gamma$ and
$\beta\approx1$.  The conservation
Eqs.~(\ref{eq:m_cons})--(\ref{eq:p_cons}) then reduce to
\begin{eqnarray}
  \label{eq:m_cons_s}
  \dot M &=& r^2 \rho u c
  \ ,\\
  \label{eq:e_cons_s}
  L \left( 1 - \frac{u}{\sigma_0^{3/2}}\right)
  &=& c\frac{(rB)^2}{4\pi}
  \ .
\end{eqnarray}
In this limit one can also simplify the expression (\ref{eq:alfspeed})
for the Alfv\'en speed in the comoving frame by using
(\ref{eq:m_cons_s}) and (\ref{eq:e_cons_s}):
\begin{equation}
  \label{eq:alf2}
  \beta_\mathrm{A}
  = \sqrt{1-\frac{u}{\sigma_0^{3/2}}}
  \quad,\qquad
  u_\mathrm{A} = \frac{\sigma_0^{3/2}}{u}
  \ .
\end{equation}

Equation~(\ref{eq:e_cons_s}) states that there is no acceleration if
$B\sim 1/r$.  This is the case in a radial outflow with ideal MHD
conditions.  We encounter here again the fact that the Poynting flux
energy in a radial ultra-relativistic MHD outflow cannot be transfered
to the matter \citep[e.g][]{begelman:94,daigne:01}.

\subsection{The evolution of the magnetic field}
\label{sec:evomag}

The evolution of the magnetic field, as it is carried with the flow,
is governed by the induction equation.  This includes the effects of
advection and field line stretching.  In addition we will include a
term to describe the decay of (a part of) the field by reconnection as
described in Sect.~\ref{sec:dissi}.  Since the reconnection is easiest
described in a local, comoving frame, we first transform the induction
equation
\begin{equation}
  \partial_t \vec B 
  = -c\, \mathbf{curl} \vec E
\end{equation}
into the comoving frame where we extend it to account for the
reconnection.

In the stationary case of our model setup the induction equation for ideal
MHD is
\begin{equation}
  \label{eq:ind}
  \partial_r \beta rB = 0
  \ .
\end{equation}
This equation describes the field evolution due to ideal MHD
processes.  To obtain the evolution term in the the comoving frame we
first need the convective derivative
\begin{equation}
  \label{eq:ind_co}
  \frac{\mathrm{d}B}{\mathrm{d}t}
  = c \beta \partial_r B
  = - \frac{cB}{r} \partial_r r\beta
\end{equation}
which then gives in terms of the comoving quantities
\begin{eqnarray}
  \label{eq:trafo0}
  \frac{\mathrm{d} B}{\mathrm{d} t}
  &=&
  \frac{\mathrm{d} (\Gamma B_\mathrm{co})}{\Gamma \mathrm{d} t_\mathrm{co}}\\
  &=&
  \frac{\mathrm{d} B_\mathrm{co}}{\mathrm{d} t_\mathrm{co}}
  + \frac{B_\mathrm{co}}{\Gamma}
  \frac{\mathrm{d} \Gamma}{\mathrm{d} t_\mathrm{co}}\\
  \label{eq:trafo}
  &=&
  \frac{\mathrm{d} B_\mathrm{co}}{\mathrm{d} t_\mathrm{co}}
  + \frac{B}{\Gamma}
  \frac{\mathrm{d} \Gamma}{\mathrm{d} t}
\end{eqnarray}
Combining (\ref{eq:ind_co}) and (\ref{eq:trafo}) gives the comoving
field evolution without dissipation effects
\begin{equation}
  \label{eq:dec_iMHD}
  \frac{\mathrm{d} B_\mathrm{co}}{\mathrm{d} t_\mathrm{co}}
  = - \frac{cB}{\Gamma r} \partial_r r u
  \ .
\end{equation}

Let us denote the striped, decayable part of the magnetic field with
$\vec B_{\rightleftharpoons}$ and the perpendicular, non-reconnecting
part with $\vec B_{\Uparrow}$.  We model the decay of the striped
component of the magnetic field in the comoving frame by the ansatz
\begin{equation}
  \label{eq:dec_dissi}
  \frac{\mathrm{d} B_{\rightleftharpoons\mathrm{co}}}
  {\mathrm{d} t_\mathrm{co}}
  = - \frac{cB_{\rightleftharpoons}}{\Gamma r} \partial_r r u
  - \frac{B_{\rightleftharpoons}}{\tau}
\end{equation}
where $\tau$ is the field decay time scale from (\ref{eq:tau}) in the
lab frame.  The non-decaying part $\vec B_{\Uparrow}$ evolves
according to the induction Eq.~(\ref{eq:ind}) so that
\begin{equation}
  \beta r B_\Uparrow = \mbox{const.}
\end{equation}
Expressing the comoving quantities in terms of lab frame quantities
similar to (\ref{eq:trafo0})--(\ref{eq:trafo}) gives
\begin{equation}
  \label{eq:trafo1}
  \frac{\mathrm{d} B_{\rightleftharpoons\mathrm{co}}}
  {\mathrm{d} t_\mathrm{co}}
  = 
  \frac{\mathrm{d} B_{\rightleftharpoons}}
  {\mathrm{d} t}
  - \frac{B_{\rightleftharpoons}}{\Gamma}
  \frac{\mathrm{d} \Gamma}{\mathrm{d} t}
  \ .
\end{equation}
Since the flow is stationary we can replace the convective derivatives
by $r$-derivatives and after combining (\ref{eq:dec_dissi}) and
(\ref{eq:trafo1}) one obtains
\begin{equation}
  c\beta \partial_r B_{\rightleftharpoons}
  - \frac{B_{\rightleftharpoons}}{\Gamma} c\beta \partial_r \Gamma
  = - \frac{cB_{\rightleftharpoons}}{\Gamma r} \partial_r r u
  - \frac{B_{\rightleftharpoons}}{\tau}
  \ .
\end{equation}
One arrives at
\begin{equation}
  \partial_r \beta rB_{\rightleftharpoons} 
  = \frac{rB_{\rightleftharpoons}}{c\tau}
\end{equation}
and with $\beta rB_{\rightleftharpoons} = \sqrt{(\beta rB)^2 - (\beta
  rB_\Uparrow)^2}$ this yields
\begin{equation}
  \partial_r \beta rB
  = - \frac{rB}{c\tau} \left[
    1 - \left(\frac{B_\Uparrow}{B}\right)^2
  \right]
  \ .
\end{equation}
Let $\mu=(B_{\Uparrow}/B)_0$ so that $\mu$ is the initial fraction
between the field strengths of the non-decaying component and the
total field.  $\mu=0$ means that the complete field decays while at
$\mu=1$ only ideal MHD processes occur and no dissipation takes place.
In the case of an equatorial outflow of an inclined rotator $\mu=\cos
i$ where $i$ is the inclination.  The other cases are more complicated
and $\mu$ is not associated to a simple geometric quantity.

Using $\mu$ as a constant of the problem the field decay equation can
be cast into this simple form:
\begin{equation}
  \label{eq:b_dec}
  \partial_r (\beta rB)^2
  = - 2 \frac{(\beta rB)^2}{\beta c\tau} \left[
    1 - \mu^2 \frac{(\beta rB)_0^2}{(\beta rB)^2}
  \right] 
  \ .
\end{equation}

Equations~(\ref{eq:m_cons}), (\ref{eq:e_cons}), (\ref{eq:p_cons}),
(\ref{eq:b_dec}) over-determine the 3 unknown functions $\rho,u,B$
because we have assumed a cold flow so that the internal energy is
neglected.  But in the ultra relativistic limit the energy and
momentum Eqs.~(\ref{eq:e_cons}), (\ref{eq:p_cons}) are equal.  In this
limit (\ref{eq:m_cons_s}), (\ref{eq:e_cons_s}), (\ref{eq:b_dec}) are
sufficient to obtain the solutions for $\rho,u,B$.

\subsection{Solution of the flow problem}

In the ultra relativistic limit $\beta\approx 1$, $\Gamma\approx u$
Eq.~(\ref{eq:b_dec}) for the evolution of the magnetic field with
distance, including dissipation becomes
\begin{equation}
  \label{eq:b_dec1}
  \partial_r (rB)^2
  = - 2 \frac{(rB)^2}{c\tau} \left[
    1 - \mu^2 \frac{(rB)_0^2}{(rB)^2}
  \right]
  \ .
\end{equation}
Using the $B$--$u$ relation (\ref{eq:e_cons_s}) to eliminate $(rB)^2$
one obtains a differential equation for $u$:
\begin{equation}
  \label{eq:udifeq}
  \partial_r u 
  = \frac{2}{c\tau} \left[
    \sigma_0^{3/2} \left( 1-\mu^2\right)
    + \sqrt{\sigma_0} \mu^2 - u
  \right]
  \ .
\end{equation}
This equation is to be integrated with initial condition
$u=u_0=\sqrt{\sigma}$.  In the absence of internal dissipation
($\mu=1$) the flow is not accelerated, $\partial_r u = 0$, as
expected.  The flow accelerates monotonically and reaches
asymptotically its terminal speed $u_\infty$ found be setting
$\partial_r u = 0$:
\begin{equation}
  \label{eq:uinf}
  u_\infty
  = \sigma_0^{3/2} \left( 1-\mu^2\right)
  + \sqrt{\sigma_0} \mu^2
  \ .
\end{equation}

The dissipation time scales are
\begin{equation}
  \tau_\mathrm{lo}
  = \frac{2\pi}{\epsilon\Omega}
  \frac{u^2}{\sqrt{1-u/\sigma_0^{3/2}}}
\end{equation}
and 
\begin{equation}
  \tau_\mathrm{tr}
  = \frac{\vartheta}{\epsilon c}
  \frac{ru}{\sqrt{1-u/\sigma_0^{3/2}}}
\end{equation}
for the longitudinal and transversal cases from (\ref{eq:tau}),
(\ref{eq:lambda_equ}), (\ref{eq:lambda_jet}), (\ref{eq:alf2}).

Since our model rests on the assumption of a significant `decayable'
component, $\mu$ in the following is taken to be of the order $0.5$
but not close to 1.  Then, the terminal velocity $u_\infty$ is much
larger than the initial velocity and $u_\infty/u_0 \approx
\sigma_0(1-\mu^2) \gg 1$ and (\ref{eq:uinf}) simplifies to
\begin{equation}
  \label{eq:uinf_sim}
  u_\infty\approx (1-\mu^2)\sigma_0^{3/2}
  \ .
\end{equation}

The differential Eq.~(\ref{eq:udifeq}) is analytically solvable at
intermediate source distances, where the flow is much faster than the
initial velocity ($\sqrt{\sigma_0}\ll u$) but is still far away from
the point where the acceleration saturates ($u \ll u_\infty <
\sigma_0^{3/2}$).  The dissipation time scales then simplify to
\begin{equation}
  \label{eq:tau_sim}
  \tau_\mathrm{lo}\approx
  \frac{2\pi u^2}{\epsilon\Omega}
  \quad,\qquad
  \tau_\mathrm{tr}\approx
  \frac{\vartheta r u}{\epsilon c}
  \ .
\end{equation}
In this case $u_\infty-u\approx u_\infty$ and (\ref{eq:udifeq})
becomes
\begin{equation}
  \partial_r u = \frac{2u_\infty}{c\tau}
\end{equation}
with the solutions
\begin{equation}
  \label{eq:uequa}
  u = \left(
    \frac{3}{\pi c} \epsilon\Omega u_\infty (r-r_0)
    + \sigma_0^{3/2}
  \right)^{1/3}
  \hfill\mbox{(long.\ case)\quad}
\end{equation}
and 
\begin{equation}
  \label{eq:ujet}
  u = \left(
    \frac{4\epsilon u_\infty}{\vartheta} 
    \ln\left(\frac{r}{r_0}\right)
    + \sigma_0
  \right)^{1/2}
  \hfill\mbox{(transversal case).\quad}
\end{equation}

\begin{figure}
  \centerline{\includegraphics{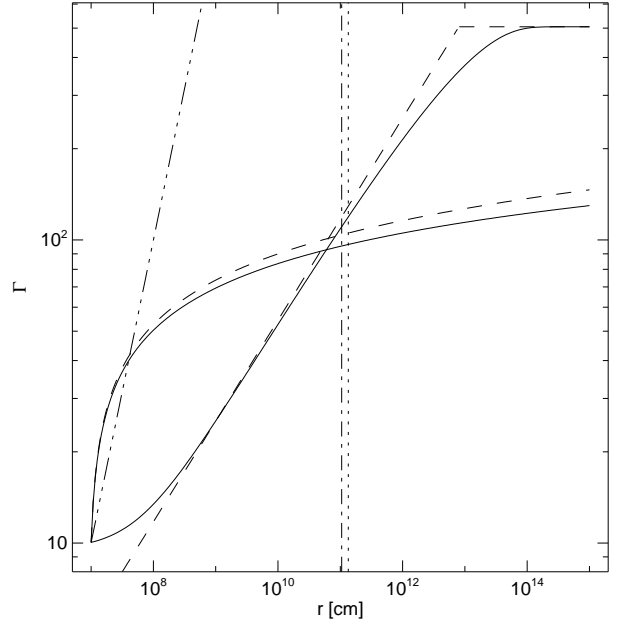}}
  \caption{Lorentz factor $\Gamma$ of the flow as 
    function of radius $r$ for the longitudinal and transversal cases.
    Model parameters are $\sigma_0=100$, $\epsilon=0.1$, $\mu^2=0.5$,
    $\Omega=10^4\,\mathrm{s}^{-1}$, $\vartheta=10^\circ$ and
    $r_0=10^7\,\mathrm{cm}$.  The solid lines are the numerical
    solutions of (\ref{eq:udifeq}) while the dashed lines are the
    approximations (\ref{eq:ujet}) and (\ref{eq:uequa_sim}).  The
    vertical lines correspond to the photospheric radii: dotted for
    the transversal case and dashed-dotted for the longitudinal case
    model.  The steep dashed-dotted line represents the $u\sim r$ law
    which is expected in the classic non-magnetic optically thick
    fireball models \protect\citep{paczynski:86}.}
  \label{fig:ucomp}
\end{figure}

The function $u$ for the longitudinal case can be described as a
broken power-law as can be seen in Fig.~\ref{fig:ucomp}.  In the
domain $r_0\ll r$, $u_0\ll u\ll u_\infty$, which we have regarded
anyway for finding the solution, the 4-velocity is well approximated
by
\begin{equation}
  \label{eq:uequa_sim}
  u = \left(
    \frac{3}{\pi c}
    \epsilon\Omega u_\infty r
  \right)^{1/3}
  \ .
\end{equation}

\subsection{The length scale for the acceleration}

In the longitudinal case the dissipation stops approximately where the
rising power-law part of the $u$ functions (\ref{eq:uequa_sim})
reaches the $u_\infty$ limit (\ref{eq:uinf_sim}).  By using
(\ref{eq:uinf_sim}) and (\ref{eq:uequa_sim}) one can write down this
\emph{saturation radius} as
\begin{eqnarray}
  r_\mathrm{sr}
  &=& \frac{\pi c}{3}
  \frac{u_\infty^2}{\epsilon\Omega}
  = \frac{\pi c}{3}
  \frac{\sigma_0^3(1-\mu^2)^2}{\epsilon\Omega}
  \nonumber\\
  \label{eq:rsr}
  &=& 7.85\cdot10^{12}\,\mathrm{cm}\cdot
  \left(\epsilon_{-1} \Omega_4\right)^{-1}
  \sigma_{0,2}^3
  \left(\frac{1-\mu^2}{0.5}\right)^2
  \ .
\end{eqnarray}
The complete analytical approximation for the longitudinal case reads
\begin{equation}
  \label{eq:uanatot}
  u = \left\{
    \begin{array}{ll}
      u_\infty \left(r/r_\mathrm{sr}\right)^{1/3}
      & \mbox{for~} r\le r_\mathrm{sr}\\
      u_\infty
      & \mbox{for~} r> r_\mathrm{sr}\\
    \end{array}
  \right.
  \ .
\end{equation}

\section{Analysis of the results}

\subsection{The photospheric radius}
\label{sec:phot}

The photosphere is located where the optical depth reaches a value of
1.  The optical depth depends on the density and the the radial
velocity $u$ and must be integrated from a finite radius to infinity.
Because we only want to know the photospheric radius within a factor
of, say 2, we define it to be where the mean free path of a photon
equals the distance from the source $r$.  In the comoving frame a
photon sees the mass density $\rho$ and the mean free path for
Thompson scattering is $1/(\kappa\rho)$.  The source distance in this
frame is $r/\Gamma$ so that the photosphere is located at
\begin{equation}
  r_\mathrm{ph}
  = \frac{u_\mathrm{ph}}{\kappa\rho_\mathrm{ph}}
\end{equation}
which yields
\begin{equation}
  \label{eq:phot}
  \frac{\kappa L}{c^3 \sigma_0^{3/2}} 
  = u_\mathrm{ph}^2 r_\mathrm{ph}
  \ .
\end{equation}

If we neglect not only the initial velocity but also the initial
radius $r_0$ compared to the photospheric radius, (\ref{eq:uequa}) and
(\ref{eq:ujet}) simplify to
\begin{equation}
  \label{eq:uphequa}
  u_\mathrm{ph}^3 =
    \frac{3}{\pi c} \epsilon\Omega (1-\mu^2)\sigma_0^{3/2} r_\mathrm{ph}
  \hfill\mbox{(longitudinal case)\ ,}\quad
\end{equation}
\begin{equation}
  \label{eq:uphjet}
  u_\mathrm{ph}^2 =
  \frac{4\epsilon (1-\mu^2)\sigma_0^{3/2}}{\vartheta} 
  \ln\left(\frac{r_\mathrm{ph}}{r_0}\right)
  \hfill\mbox{(transversal case)\ .}\quad
\end{equation}
Together with condition (\ref{eq:phot}) at the photosphere we arrive
at the equations for the photospheric radius and the 4-velocity in the
longitudinal case
\begin{eqnarray}
  u_\mathrm{ph}
  &=& \left[
    \frac{3\kappa}{\pi c^4} \epsilon\Omega (1-\mu^2)L
  \right]^{1/5}\nonumber\\
  &=& 119\cdot \left[
    \epsilon_{-1} \Omega_4
    \left(\frac{1-\mu^2}{0.5}\right) L_{50}
  \right]^{1/5} \ ,
  \label{eq:uph_sim}\\
  r_\mathrm{ph}
  &=& \left[
    \frac{\pi^2 \kappa^3}{9 c^7}
    \frac{L^3}{\left(\epsilon\Omega(1-\mu^2)\right)^2}
  \right]^{1/5}
  \sigma_0^{-3/2}\nonumber\\
  &=& 1.05\cdot10^{11}\,\mathrm{cm}\nonumber\\
  && \quad\cdot \left[
    \epsilon_{-1} \Omega_4 \left(\frac{1-\mu^2}{0.5}\right)
  \right]^{-2/5}
  L_{50}^{3/5}
  \sigma_{0,2}^{-3/2}
  \ .  
  \label{eq:rph_sim}
\end{eqnarray}
Note that the 4-velocity at the photosphere $u_\mathrm{ph}$ does not
depend on the initial Poynting flux ratio $\sigma_0$ and only weakly
on $L$.

For the transversal case the flow velocity always depends greatly on
the initial radius $r_0$.  The dissipation time scale is
$\tau_\mathrm{tr}\sim ru$ and most energy is released at small $r$
near the source.  The acceleration depends crucially on the onset of
the dissipation and therefore on $r_0$.  In our simple model $r_0$ and
$\vartheta$ are not well determined by physical arguments so that the
transversal case is rather uncertain and highly speculative.  One
cannot write down robust equations for the photosphere like in the
longitudinal case without many degrees of freedom.

\subsection{Energy available for prompt radiation}

The energy dissipated beyond the photospheric radius is
\begin{equation}
  L_\mathrm{D} 
  = \left(u_\infty - u_\mathrm{ph}\right) \dot M c^2
  \ .
\end{equation}
Using (\ref{eq:Mdot}), (\ref{eq:uinf_sim}) and (\ref{eq:uph_sim}) this
yields
\begin{equation}
  L_\mathrm{D}
  = e \left(1-\mu^2\right) L  
\end{equation}
with
\begin{equation}
  \label{eq:eff}
  e = 1 - \left(
    \frac{3\kappa}{\pi c^4}
    \frac{\epsilon\Omega L}{\left(1-\mu^2\right)^4}
  \right)^{1/5} \sigma_0^{-3/2}
  \ .
\end{equation}
For a Poynting flux dominated flow, the magnetic energy flux equals
the total energy flux $L$.  Of this, a fraction $1-\mu^2$ is
dissipated internally while a fraction $e(1-\mu^2)$ can be converted
to radiation beyond the photosphere.  Thus $e$ is an \emph{efficiency
  factor}, which gives the ratio between the energy dissipated in the
optically thin domain and the total dissipated energy.  Efficient
conversion of free magnetic energy into non-thermal radiation can
happen if $e$ is of the order unity which requires that the second
term in (\ref{eq:eff}) is small:
\begin{equation}
  \label{eq:eff1}
  0.24\cdot
  \left(
    \epsilon_{-1} \Omega_4 L_{50}
  \right)^{1/5}
  \left(\frac{1-\mu^2}{0.5}\right)^{-4/5}
  \sigma_{0,2}^{-3/2}
  < 1
\end{equation}
or written differently
\begin{equation}
  \label{eq:eff2}
  \sigma_0
  >
  39\cdot
  \left(
    \epsilon_{-1} \Omega_4 L_{50}
  \right)^{2/15}
  \left(\frac{1-\mu^2}{0.5}\right)^{-8/15}
\end{equation}
where $\epsilon_{-1} = \epsilon/0.1$, $\Omega_4 =
\Omega/\left(10^4\,\mathrm{s^{-1}}\right)$, $L_{50} =
L/\left(10^{50}\,\mathrm{erg\,s^{-1}\,sterad^{-1}}\right)$ and
$\sigma_{0,2} = \sigma_0/100$ are parameters scaled to fiducial GRB
values.

When (\ref{eq:eff2}) is satisfied, part of the magnetic energy is
released beyond the photosphere, and powers the prompt radiation.  If
it is not satisfied, the energy is released inside the photosphere and
is converted, instead, into bulk kinetic energy.  Some other means of
conversation into radiation is then needed, such as internal shocks.
Since the dependence on parameters other than the initial Poynting
flux ratio $\sigma_0$ is small in (\ref{eq:eff2}), we conclude that
efficient powering of prompt radiation by magnetic dissipation in GRB
is possible for $\sigma_0\ga 100$.

\subsection{Longitudinal and transversal cases in comparison}

The major difference between the longitudinal and the transversal case
is the different dissipation time scale.  While the decay time scale
for the longitudinal case (\ref{eq:tau_sim}) is $\tau_\mathrm{lo}\sim
u^2$ and therefore limited by $u\le u_\infty$, the time scale for the
transversal case $\tau_\mathrm{tr}\sim ru$ is not limited.  At small
radii it starts at low values but grows then to infinity.  This major
difference is visualised in Fig.~\ref{fig:ucomp} where the flow
Lorentz factor is plotted depending on the radius.

\begin{figure}
  \centerline{\includegraphics{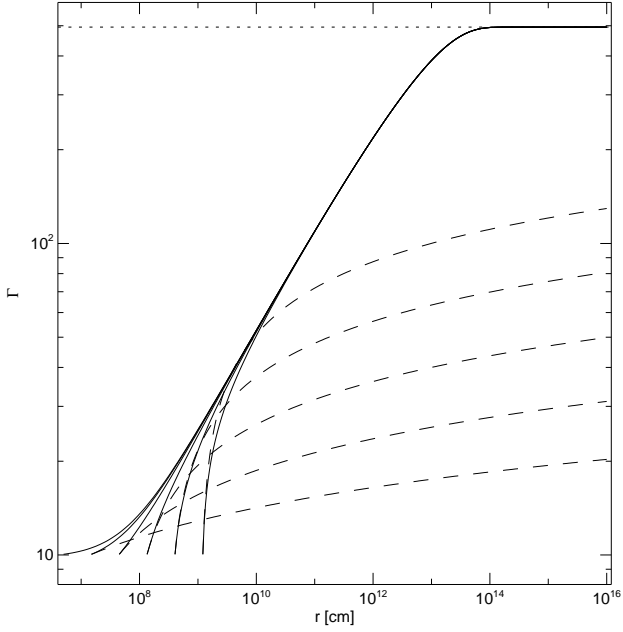}}
  \caption{The influence of $r_0$ on the Lorentz factor: the solid 
    lines correspond to longitudinal case solutions and the dashed one
    to transversal case solutions with $1.5\cdot10^{7}\,\mathrm{cm}\le
    r_0\le 1.2\cdot10^9\,\mathrm{cm}$.  The constants used where
    $\sigma_0=100$, $\mu^2=0.5$, $\epsilon=0.1$,
    $\Omega=10^4\,\mathrm{s}^{-1}$.  For the transversal cases
    $\vartheta=2\pi c\sqrt{\sigma_0}/(\Omega r_0)$ is chosen so that
    the initial acceleration (slope at $r_0$) is the same as in the
    corresponding longitudinal cases.}
  \label{fig:r0-dependence}
\end{figure}
As mentioned in Sect.~\ref{sec:phot} the transversal case depends
strongly on the initial radius $r_0$.  This is seen in
Fig.~\ref{fig:r0-dependence} where the numerical solutions of
(\ref{eq:udifeq}) are shown for various initial radii.  While all
longitudinal case solutions merge toward the $u\sim r^{1/3}$ power-law
there is a large spread in the transversal case solutions.

The longitudinal and transversal cases are the two limits for a
general case where both kinds of scaling of the decay time scale
occur.  One can model the mixing of both cases by writing the
dissipation time scale as
\begin{equation}
  \tau = k \left(\frac{r}{r_0}\right)^\alpha
  \left(\frac{u}{u_0}\right)^{2-\alpha}
\end{equation}
where $0<\alpha<1$ is a dimensionless parameter which determines the
mixing.  $\alpha=0$ corresponds to the pure longitudinal case and
$\alpha=1$ to the transversal case.  The constant $k$ can be written
depending on the corresponding model parameters as
\begin{equation}
  k = \frac{2\pi\sigma_0}{\epsilon\Omega}  
\end{equation}
or as 
\begin{equation}
  k = \frac{\vartheta r_0 \sqrt{\sigma_0}}{\epsilon c}
  \ .
\end{equation}

\begin{figure}
  \centerline{\includegraphics{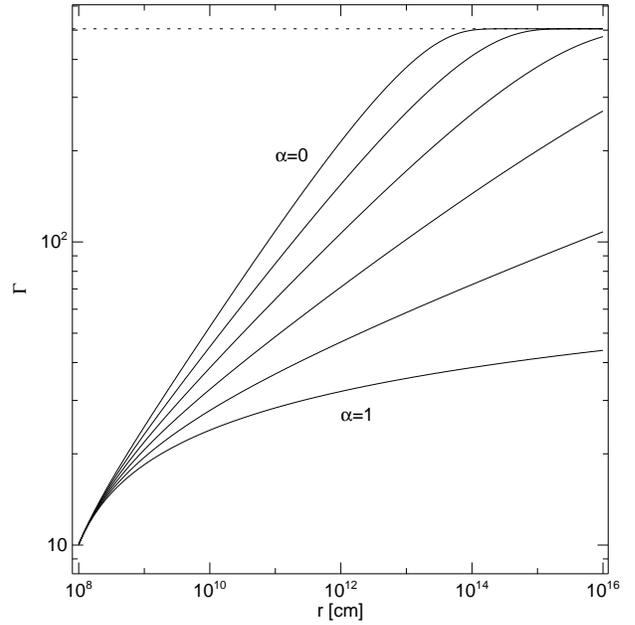}}
  \caption{The Lorentz factor $\Gamma$ for different parameters 
    $\alpha\in\{0,0.2,0.4,0.6,0.8,1\}$.  The other wind parameters
    were set to $\sigma_0=100$, $\mu^2=0.5$, $\epsilon=0.1$,
    $r_0=10^8\,\mathrm{cm}$, $\Omega=10^4\,\mathrm{s}^{-1}$
    (corresponding to $\vartheta=10.8^\circ$ in the transversal
    description).  The different winds all start with the same
    dissipation rate so that they show the same initial acceleration.
    $\Gamma_\infty$ is marked by a horizontally dotted line.}
  \label{fig:alpha}
\end{figure}
Figure~\ref{fig:alpha} shows the velocity profiles for various
$\alpha$ values.  All graphs result from a numerical integration of
(\ref{eq:udifeq}).  Beyond the photosphere, assuming it is around
$10^{11}\,\mathrm{cm}$, the dissipation is only efficient if the field
variation does not point in transversal direction.  The efficiency
estimation in (\ref{eq:eff}) is therefore an upper limit for the
general case with $\alpha\not=0$.

\subsection{The validity of the MHD condition}
\label{sec:mhd}

Because we work with the ideal MHD approximation we have to make sure
that there are enough charges in the flow to make up the required
electric current density.  Because the reconnection processes will
destroy the ordered initial field configuration quickly it does not
make much sense to consider this configuration throughout the flow.
But one can at least estimate needed currents by looking at a
sinusoidal wave in the equatorial plane.  In \citetalias{spruit:01} we
derived the limiting radius where the MHD condition breaks down by
using a constant flow speed and assumed $\mu=0$.  The condition that
enough charges are available to carry the current is
\begin{equation}
  \frac{\Omega B}{4\pi\Gamma^2}
  = \frac{2\rho ec}{m_\mathrm{p}}
  \ .
\end{equation}
This yields the radius up to which the MHD approximation holds:
\begin{eqnarray}
  \label{eq:rMHD}
  r_\mathrm{MHD} 
  &=& \frac{4e \sqrt{\pi L}}{m_\mathrm{p} c^{3/2} \Omega}
  \frac{u/\sigma_0^{3/2}}{\sqrt{1-u/\sigma_0^{3/2}}}\\
  \label{eq:rMHD1}
  &>& 4\cdot10^{18}\,\mathrm{cm}\cdot
  L_{50}^{1/2} \Omega_4^{-1} \sigma_{0,2}^{-1}
  \ .
\end{eqnarray}
Here, we have used the dependence (\ref{eq:e_cons_s}) of the magnetic
field strength on velocity.  In (\ref{eq:rMHD}) $r_\mathrm{MHD}$ is
written as a function of $u$ and depends implicitly on $r$.  At $r_0$
where $u=u_0=\sqrt{\sigma_0}$ it starts at the value (\ref{eq:rMHD1})
and rises strongly until the final velocity $u = u_\infty =
\sigma_0^{3/2}$ is reached where $r_\mathrm{MHD}$ diverge to $\infty$.
For the GRB parameters assumed here, we find $r_\mathrm{MHD}\gg
r_\mathrm{sr}$ and the MHD approximation is always fulfilled, as in
\citetalias{spruit:01}.

\subsection{Comparison with the striped pulsar wind}
\label{sec:pulsars}

Dissipation of magnetic energy was applied to the Crab pulsar wind by
\citet{lyubarsky:01}.  Their model setup included a striped pulsar
wind \citep{coroniti:90} that is the equatorial wind of on inclined
rotator.  This is quite similar to our longitudinal case where all
Poynting flux can decay so that $\mu=0$.  The wind starts with
$\Gamma=\sqrt{\sigma_0}$ and reaches $\Gamma_\infty = \sigma_0^{3/2}$
as in our model.  Due to a difference approach to model the
reconnection rate they obtain a flow acceleration of $\Gamma \sim
r^{1/2}$ \citep[ Eq.~(30)]{lyubarsky:01} which is faster than
$\Gamma\approx u \sim r^{1/3}$ form (\ref{eq:uanatot}).

The findings of \citet{lyubarsky:01} that the reconnection is
inefficient for the Crab wind seems to contradict our result, that it
efficiently accelerates the GRB outflow.  The reason for that is the
different initial Poynting flux values used for the Crab pulsar and in
our study.  $\sigma_0$ is the critical parameter controlling the final
Lorentz factor and the spatial size of the accelerating wind.

Discussing the Crab pulsar wind in detail and speculating why
reconnection fails is beyond the scope of the present paper.  Instead,
we simply take the flow parameter values $\sigma_{0,2}=400$,
$\Omega_4=0.02$ from \citet{lyubarsky:01} and show that our model
gives basically the same result as the striped wind model.  However,
see \citet{lyubarsky:01a} for a critical revision of the Crab pulsar
wind parameters.  Equation~(\ref{eq:uanatot}) yields for the
4-velocity at the observed termination shock at
$r=3\cdot10^{17}\,\mathrm{cm}$
\begin{equation}
  \frac{u}{u_\infty}
  = 0.023\,\epsilon_{-1}^{-1/3}
  \quad,\qquad
  \frac{u}{u_0}
  = 920\,\epsilon_{-1}^{-1/3}\ .
\end{equation}
When the wind reaches its termination shock only a small fraction of
the Poynting flux was converted.  Though, due to the small amount of
mass in the flow large acceleration occurs and the Lorentz factor
increases by almost 4 orders of magnitude.  This is the same result as
obtained by \citet{lyubarsky:01}.  The different acceleration laws of
the two models does not change the picture.  The observed pulsar wind
bubble is to small to allow for efficient reconnection.

The radius $r_\mathrm{sr}$ from (\ref{eq:rsr}) denotes the radius
where the Poynting flux conversion ends.  Its value scales with the
third power of $\sigma_0$.  Plausible Lorentz factors for GRB winds of
around $10^2$--$10^4$ imply $\sigma_{0,2} \approx 0.2$--$5$ (or larger
for $\mu>0$).  This lowers $r_\mathrm{sr}$ by 6 orders of magnitudes
compared to the Crab wind.  Thus $r_\mathrm{sr} \la
3\cdot10^{15}\,\mathrm{cm}$ which is smaller than the radius
$r_\mathrm{a} \approx 10^{16}\,\mathrm{cm}$ where the flow runs into
the ambient medium \citep{piran:99}.
The requirement on $\sigma_0$ for the dissipation to take place inside
a radius $r_\mathrm{a}$ can be expressed by (\ref{eq:rsr}) which
yields an upper limit for the initial Poynting flux ratio:
\begin{equation}
  \sigma_0 \la 
  1100 \cdot
  \left( r_{\mathrm{a},16} \epsilon_{-1} \Omega_4\right)^{1/3}
  \left(\frac{1-\mu^2}{0.5}\right)^{-2/3}
  \ .
\end{equation}
For GRBs there is no size problem as for the Crab wind and Poynting
flux can be efficiently converted.

\section{Discussion}
\label{sec:disc}

We have investigated the effect of dissipation of magnetic energy in a
GRB outflow on the acceleration of the flow.  Such dissipation is
expected if the flow contains small scale changes of direction of the
field for example when the flow is produced by the the rotation of a
non-axisymmetric magnetic field.  The dissipation is governed by the
speed of fast reconnection, parameterised in our calculations as a
fraction $\epsilon\approx 0.1$ of the local Alfv\'en speed in the
flow.

Two possibilities for the field geometry in the outflow have been
considered: a geometry where the changes in the small scale field
direction occur along the bulk flow direction, and a geometry where
the field variation is transversal to the flow direction.  The first
mentioned, \emph{longitudinal case} is expected in the equatorial
plane of an inclined rotator as in the `striped' pulsar wind model of
\citet{coroniti:90}.  The second, \emph{transversal case} can be
associated with a polar outflow where the field line structure
resembles a spiral.  In both cases there are MHD instabilities
(tearing and kink instabilities) which lead to reconnection processes.
They differ only by the functional form of the reconnection time
scale.

We find that in any case the process leads to a strong increase of the
bulk Lorentz factor of the flow.  This acceleration is due to the
outward decrease of the magnetic pressure resulting from the field
decay.  At the same time, the dissipated energy can be released to
large extend in the optically thin part of the flow beyond its
photosphere, and can power most if not all of the prompt emission.
This provides an alternative to the internal shock model.

The calculation is done for a stationary wind.  Why this approximation
is valid for highly variable objects like GRBs is not obvious.  The
duration of GRBs $t$ is of the order of a few seconds.  One can
approximate the wind as stationary within a source distance
$ct\approx10^{11}\,\mathrm{cm}$.  Thus the flow up to the photospheric
radius is well described by a stationary description.  Further out the
time dependence of a real flow will become more important but that
topic is beyond the scope of this work.

The outflow with transversal field variation contains some additional
complications which does not occur in the longitudinal case.  The
dissipation time scale is proportional to the source distance.  This
results in a rapid energy dissipation near to the source and the
velocity profile depends critically on the radius where the
dissipation sets in.  But this initial radius is hard to estimate from
first principles.

We have used the spiral-like field geometry of a polar flow as
pictured in \citetalias{spruit:01} to justify the existence of
transversal field variations.  This field geometry occurs for a polar
outflow of an axisymmetric rotator.  The following arguments give
reasons why this field geometry is rather special and may not be
important in a general.  The kink instability leads to a break-down of
the ordered spiral field configuration.  After some Alfv\'en crossing
times the field geometry will have changed so that the `longitudinal'
dissipation time will become important while the `transversal' time
scale grows large and can be neglected.  On the other hand the rotator
may not be perfectly aligned and non-axisymmetric field components are
also present in the polar outflow.  So, we probably have always
longitudinal field variations in the flow so that the findings found
in our treatment of the `longitudinal case' might be much more
applicable and general.

We assume that the thermal energy flux is negligible compared to the
kinetic and Poynting energy flux.  The temperature is set to zero
which simplifies the treatment and allows an analytical integration of
the dynamic equations.  Setting the thermal pressure gradient
artificially to zero might appear to underestimate the acceleration.
On the other hand the energy equation takes care that all released
magnetic energy shows up in kinetic form.  In fact, we overestimate
the acceleration by doing so because the energy part converted into
heat reduces the the gain of kinetic energy in the flow.  Another
physical argument explains why the flow stays cold: The acceleration
expressed by the scaling of the Lorentz factor gives $\Gamma\sim
r^{1/3}$ for our model.  The release of magnetic energy must therefore
also scale with $r^{1/3}$.  In contrast to that, purely thermal
acceleration by adiabatic cooling leads to more rapid flow
acceleration where the Lorentz factor scales like $\Gamma\propto r$
\citep{paczynski:86}.  Thus, heating proceeds slower than adiabatic
cooling so that the thermal pressure gradient is not important
compared to the magnetic pressure gradient which drives the flow.  The
reason why \citet{lyubarsky:01} find a faster acceleration of
$\Gamma\sim r^{1/2}$ in a similar model lies in the different
reconnection prescription and is not due to their inclusion of thermal
pressure.

In the optically thin regime part of the dissipated energy radiates
away.  There, the model over-estimates the gain of kinetic energy.  We
cannot give arguments how much dissipated energy escapes as prompt
radiation so that the total amount of released energy gives only an
upper limit on the Lorentz factor.

The photospheric radius determines the lower limit on radius for the
region in which non-thermal radiation is expected to originate.  For
typical GRB parameters describing the total luminosity, the baryon
loading, the fraction of dissipatable energy and the reconnection rate
one finds that a considerable amount of dissipation takes place in the
optically thin region.  Part of the dissipated energy is converted
into non-thermal radiation.  The remainder still leads to an
acceleration of the flow.  This acceleration is caused by the magnetic
pressure gradient induced by the field dissipation.  Since the
acceleration continues outside the photosphere up to the radius where
all the free magnetic energy is used up this non-thermal radiation is
emitted from matter with different Lorentz factors.  The observable
spectrum in thus smeared out compared to a spectrum from a uniformly
moving medium.  For a more sound analysis of this topic one needs a
model for the radiation process.

The Poynting flux conversion happens at radii $r\la
10^{15}\,\mathrm{cm}$ which is inside the distance
$\approx10^{16}\,\mathrm{cm}$ where the GRB outflow is expected to run
into the external medium.  Thus, the Poynting flux can be converted
efficiently.  But by applying the model to the Crab pulsar wind we
come to the same conclusions as \citet{lyubarsky:01}: The conversion
is inefficient since the observed pulsar wind bubble is to small to
contain the whole region where reconnection takes place.  For the Crab
pulsar the assumed initial Poynting flux ratio is larger than for GRBs
leading to a much longer reconnection phase.  The presented model does
not settle this Crab wind problem.

The most important parameter which controls the amount of energy
dissipated beyond the photosphere is the initial Poynting flux to
kinetic energy flux ratio.  If its value is around 100 or greater much
non-thermal, prompt emission is produced.  If its value is of the
order of 10, however, all the Poynting flux energy is converted into
kinetic energy and thermal radiation.  Only prompt thermal emission
and afterglow emission is expected in this case.  The initial Poynting
flux ratio is a measure for the baryon loading in a sense that a high
baryon loading corresponds to a low initial Poynting flux ratio.
Observations indicate that X-ray flashes and X-ray rich GRBs are very
similar phenomena which probably differ only by the amount of baryon
loading \citep{heise:01}.  In the context of our model, X-ray flashes
can be associated with low initial Poynting flux ratios.  In this
case, the X-ray emission is thermal radiation from the photosphere.
Increasing the initial Poynting flux ratio leads to the emission of
non-thermal $\gamma$-rays in the optically thin region, thus producing
X-ray rich and regular GRBs.  If afterglows of X-ray flashes could be
observed they would yield information about the connection to regular
GRBs.  Afterglows depend less strongly on the initial Poynting flux
ratio but rather on the total luminosity of the outflow.  Thus, X-ray
flash afterglows should be similar to afterglows of regular GRBs
according to our model.  In a future work we will investigate the
thermal emission more quantitatively.

The model predicts black-body radiation originating from the
photosphere of the flow.  We can calculate the radius of the
photosphere and the Lorentz factor of the flow there.  Together with
the temperature one is able to calculate the luminosity if the thermal
radiation.  Since our approximation treats the flow as cold we cannot
give quantitative results in this respect.  Though, one finds that the
Lorentz factor at the photosphere depends only weakly on the model
parameters.  Therefore, the observable temperature $kT_\mathrm{obs} =
\Gamma_\mathrm{ph} kT/(1+z)$ of the thermal component of a GRB depends
primarily on the redshift $z$ and the temperature in the comoving
frame $T$.  This result simplifies the task to disentangle the effects
of different model parameters on the temperature.  A detailed,
quantitative analysis of the thermal radiation will be done in a
following study.

\begin{acknowledgements}
  I thank H.C.~Spruit for enlightening discussions and the critical
  reading of the manuscript.
\end{acknowledgements}

\bibliographystyle{aa}
\small
\bibliography{h3387}

\end{document}